\documentclass[english]{article}
\usepackage[T1]{fontenc}
\usepackage[latin9]{inputenc}
\usepackage{geometry}
\usepackage{color}
\usepackage{babel}
\usepackage{amsmath}
\usepackage{amsthm}
\usepackage{setspace}
\usepackage[unicode=true,pdfusetitle,
 bookmarks=true,bookmarksnumbered=false,bookmarksopen=false,
 breaklinks=false,pdfborder={0 0 1},backref=false,colorlinks=false]
 {hyperref}

\makeatletter
  \theoremstyle{plain}
  \newtheorem{conjecture}{\protect\conjecturename}

\makeatother

  \providecommand{\conjecturename}{Conjecture}

\begin{document}
\begin{doublespace}
\begin{center}
\textbf{\Large{}Solving Society\textquoteright s Big Ills, A Small
Step}
\par\end{center}{\Large \par}

\begin{center}
\textbf{Ravi Kashyap }
\par\end{center}

\begin{center}
\textbf{City University of Hong Kong}
\par\end{center}

\begin{center}
\begin{center}
\today
\par\end{center}
\par\end{center}

\begin{center}
\par\end{center}

\begin{center}
\textbf{\textcolor{blue}{\href{http://www.ccsenet.org/journal/index.php/ass/article/view/66369}{Edited Version: Kashyap, R. (2017). Solving Society\textquoteright s Big Ills, A Small Step. Asian Social Science, 13(4), 175-191. }}}\tableofcontents{}
\par\end{center}
\end{doublespace}

\begin{doublespace}

\section{Abstract }
\end{doublespace}

\begin{doublespace}
We look at a collection of conjectures with the unifying message that
smaller social systems, tend to be less complex and can be aligned
better, towards fulfilling their intended objectives. We touch upon
a framework, referred to as the four pronged approach that can aid
the analysis of social systems. The four prongs are:
\end{doublespace}
\begin{enumerate}
\begin{doublespace}
\item The Uncertainty Principle of the Social Sciences
\item The Objectives of a Social System or the Responsibilities of the Players 
\item The Need for Smaller Organizations 
\item Redirecting Unintended Outcomes\end{doublespace}

\end{enumerate}
\begin{doublespace}
Smaller organizations mitigating the disruptive effects of corruption
is discussed and also the need for organizations, whose objective
is to foster the development of other smaller organizations. We consider
a way of life, which is about respect for knowledge and a desire to
seek it. Knowledge can help eradicate ignorance, but the accumulation
of knowledge can lead to overconfidence. Hence it becomes important
to instill an attitude that does not knowledge too seriously, along
with the thirst for knowledge. All of this is important to create
an environment that is conducive for smaller organizations and can
be viewed as a natural extension of studies that fall under the wider
category of understanding factors and policies aimed at increasing
the welfare or well-being to society.
\end{doublespace}

\begin{doublespace}

\section{Introduction }
\end{doublespace}

\begin{doublespace}
\textquotedblleft We human beings are social beings. We come into
the world as the result of others\textquoteright{} actions. We survive
here in dependence on others. Whether we like it or not, there is
hardly a moment of our lives when we do not benefit from others\textquoteright{}
activities. For this reason, it is hardly surprising that most of
our happiness arises in the context of our relationships with others.\textquotedblright{}
- Dalai Lama XIV 

\textquotedblleft Human beings are social creatures. We are social
not just in the trivial sense that we like company, and not just in
the obvious sense that we each depend on others. We are social in
a more elemental way: simply to exist as a normal human being requires
interaction with other people.\textquotedblright{} - Atul Gawande 

Society and the problems facing society then simply become manifestation
of the interactions between human beings. These interactions are complex
and the complexity increases, the greater the scope of what a social
system is trying to accomplish. The greater the number of uncertain
human actions that can impact a system, the greater will be the complexity
and uncertainty of that social system. The more the participants in
a system, the more will be the number of actions and hence the human
uncertainty element can be reduced by reducing, the number of humans
that can impact that social system. While, natural sources of uncertainty
certainly abound and can be extremely impactful, a detailed discussion
of those will be postponed for another time.

(Bacharach 1989) sets down a set of ground rules and vocabulary to
facilitate focused discussion about the structure of organization
and management theories. A matrix of criteria for evaluating the variables,
constructs, and relationships that together compose a theory is developed,
while distinguishing theory from mere description. The primary goal
of a theory is to answer the questions of how, when, and why, unlike
the goal of description, which is to answer the question of what. 

(Whetten 1989) also goes into the What, How, Why, Who, Where and Whens
of a good theory. Any theory theory may be evaluated based on (a)
falsifiability and (b) utility. Falsifiability determines whether
a theory is constructed such that empirical refutation is possible.
Utility refers to the usefulness of theoretical systems. A theory
is useful if it can both explain and predict. (Nagel, 1961; Popper,
1959) have a set of rules for the examination of the constructs and
variables which are the units of theoretical statements.

First we start with a conjecture that describes, why we need the main
contribution of this paper, a technique for the analysis of social
systems, termed, the Four Pronged Approach. As the paper unfolds,
it will become clear that the four pronged approach attempts to provide
all the elements that any good theory requires and lays down a conceptual
framework that can be used to assess the effectiveness of the main
points. 

The application of this methodology leads to the main proposition
of this paper that organizations are becoming more complex and bad
outcomes are happening despite the good intentions of the participants
involved. The application of this technique provides the conclusion
that smaller organization size is the answer to many current social
problems. We present a collection of conjectures that illustrate the
issues with large organizations and how smaller systems can better
cope with these issues. We briefly touch upon ways in which we can
move towards smaller systems.
\end{doublespace}
\begin{conjecture}
\begin{doublespace}
\textbf{The beauty and also the bane of measurement is that there
is always something better or bigger and there is also something worse
or smaller. }\end{doublespace}

\end{conjecture}
\begin{doublespace}
This follows partly from the current practice of mapping anything
that can be measured to real numbers which are infinite and noting
that the universe seems to be infinite and the smallest particle has
not been discovered and it has not been proved that there is nothing
smaller. This means that there is no absolutely satisfactory answer
to the question, how small is small enough? What we can hope to accomplish
is establish a relative scale that can show what an optimal size might
be for a particular purpose, by looking look at smaller systems and
comparing them to larger systems with similar objectives, but yielding
dissimilar results. The four pronged approach is one such analysis
technique.

We need to consider all the four prongs because the first one tells
us about the limitations of any relationships we uncover; the second
tells us about the overriding need of any social system and aids in
the verification of whether we are deviating from the intended goals;
the third tells us that keeping complexity in check is important for
accomplishing our objectives. One way for doing that is to have smaller
systems with limited scope; and the fourth one tells us where unintended
outcomes, that provide no real benefit, can result, despite the care
we take to adhere to the stipulations of the first three. Just like
the four directions, we need to be aware that there is a degree of
interconnectedness in the below four prongs.

1. The Uncertainty Principle of the Social Sciences 

2. The Objectives of a Social System or the Responsibilities of the
Players

3. The Need for Smaller Organizations

4. Redirecting Unintended Outcomes
\end{doublespace}
\begin{conjecture}
\begin{doublespace}
\textbf{The four prongs are like the four directions for an army general
looking for victory and any attempt at reform that does not consider
all the four prongs will prove to be insufficient and will be incomplete
at best.}\end{doublespace}

\end{conjecture}
\begin{doublespace}
The rest of the discussion will substantiate the above claim and its
applicability to study social science phenomenon. It is worthwhile
to mention here that for most assertions made below, numerous counter
examples and alternate hypothesis can be produced. These are strictly
attempts at tracing the essentials rather than getting bogged down
with a specific instance. However, any study requires forming a conceptual
framework based on the more common observations, yet being highly
attuned to any specifics that can stray from the usual. Also, for
the sake of brevity, a number of finer points have been omitted and
certain simplifying assumptions have been made. Given the scope and
complications of the below discussion, drawbacks are hard to avoid
and future iterations will seek to address these as they are discovered.
\end{doublespace}

\begin{doublespace}

\section{The Four Pronged Approach }
\end{doublespace}

\begin{doublespace}

\subsection{The Uncertainty Principle of the Social Sciences}
\end{doublespace}

\begin{doublespace}
Kashyap (2014a) discuss the uncertainty principle of the social sciences,
though parts of the discussion are generic, the greater portion centers
around, the themes of investing and financial services, paving the
way for this present work to consider a broader setting.
\end{doublespace}
\begin{conjecture}
\begin{doublespace}
\textbf{The Uncertainty Principle of the Social Sciences can be stated
as, \textquotedblleft Any generalization in the social sciences cannot
be both popular and continue to yield predictions, or in other words,
the more popular a particular generalization, the less accurate will
be the predictions it yields\textquotedblright .} 

\textup{This is because as soon as any generalization and its set
of conditions becomes common knowledge, the entry of many participants
shifts the equilibrium or the dynamics, such that the generalization
no longer applies to the known set of conditions. An observation is
likely to be more popular when there are more people comprising that
system; and it is important to try and explicitly understand, where
possible, how predictions can go awry. Every social system then operates
under the overarching reach of this principle.}\end{doublespace}

\end{conjecture}
\begin{doublespace}

\subsection{The Objectives of a Social System or the Responsibilities of the
Players}
\end{doublespace}

\begin{doublespace}
The ultimate objective of any social system is to maximize well-being.
To determine what is intrinsic to well-being, requires acknowledging
its subjectivity. Layard (2010) talks about surveys in the United
States that showed no increase in happiness over the past 60 years,
reflecting the fact that higher national income has not brought the
better quality of life that many expected. The science of subjective
well-being is young, but it is developed enough to know that we need
to collect data and make it a prime objective to quantitatively study
the determinants of well-being, so that it can be used in policy analysis.
\end{doublespace}
\begin{conjecture}
\begin{doublespace}
\textbf{Poverty is a state of mind}. \end{doublespace}

\end{conjecture}
\begin{doublespace}
It is important to gain a more profound, comprehension of welfare
and delineate its components into those that result from an increase
in goods and services, and hence can be attributed to economic growth,
and into those that are not related to economic growth but lead to
a better quality of life. The reasoning here being that economic growth
alone is an inadequate indicator of well-being. Hand in hand with
a better understanding of the characteristics of welfare, comes the
need to consider the measures or metrics we currently have that gauge
economic growth and supplement those with factors that capture well-being
more holistically. This is important because, there would be little
sense in pursuing policies aimed at increasing some widely used metric
like Gross Domestic Product, GDP, if such policies do not lead to
an increase in welfare and worse still, if they lead to an unintentional
decrease in well-being; on a lighter note, it is worth pondering about
which meaning of gross is applicable in the context of GDP.

There is a compelling case for constructing better metrics to measure
welfare. Stiglitz, Sen \& Fitoussi (2009) highlighted the deficiencies
in existing metrics, encapsulated an agenda for improvements, and
discussed key areas on which further research is needed. An alternative
to GDP can use variables that show the increase in essentials like
food, health care, education, real estate prices, and disposable income.
We need metrics that capture not just the increase, but measure the
distribution of consumption goods and supplement those with ways that
gauge how quality of life improves. 

A proxy for quality of life can be captured with variables for: environmental
factors relating to air, water, noise pollution; leisure time per
day; vacation time; personal safety against crimes and conflicts;
social factors like availability of support in case of need; freedom
to express oneself and political participation; political stability;
and lastly, sustainable ways of production, or having a sustainability
index based on depletion of natural resources, which boils down to
making sure that what we produce today can continue to be produced
with minimal impact to the environment and being able to maintain
the current level of well-being for future generations. (Kahneman
and Krueger 2006) emphasize the importance of subjective well-being
and argue that it is fruitful to distinguish among different conceptions
of utility rather than presume to measure a single, unifying concept
that motivates all human choices and registers all relevant feelings
and experiences.
\end{doublespace}
\begin{conjecture}
\begin{doublespace}
\textbf{The question of what is absolutely imperative to lead a good
life is a constantly changing one, as luxuries end up becoming necessities. }\end{doublespace}

\end{conjecture}
\begin{doublespace}
This tells us that there can be no absolute measure of well-being.
We need to be mindful of the limitations of any static measure and
include variables that capture the change in our consumption habits
over the years, especially across the constituents of any welfare
measure. As we look to analyze the objectives of a social system,
we should look beyond measures of how this system contributes to the
GDP, but also consider other variables that provide a more complete
picture of how this system contributes to welfare.

Social systems can be broadly categorized into those that are profit
seeking, we shall refer to them as businesses, and the others that
are eleemosynary. We consider the case of the business or the profit-seeking
corporation. Friedman (1970) suggests that the social responsibility
of business is to increase profits. Recognizing that attributing objectives
to any social system is a vague concept, it becomes important to ask
precisely, what this implies for whom. The discussion then moves on
to the responsibilities of corporate executives, who are expected
to make profit maximizing decisions while conforming to the rules
of society, embodied in law and ethical custom and stay within the
rules of the game. In order to fulfill such an expectation,
\end{doublespace}
\begin{conjecture}
\begin{doublespace}
\textbf{It is implicitly assumed that the corporate executive is a
philosopher king, a concept dating back to Plato (The republic of
Plato), wise enough to know what is right, with the authority to enforce
it and the self-control to not abuse his power}. 

\textup{There is an additional assumption here that the shareholders,
whose agent is the executive, are capable of selecting the right person,
of monitoring his behavior and taking corrective measures. Surely,
that is a tall order in the complex business environment that is today.}\end{doublespace}

\end{conjecture}
\begin{doublespace}
The issue of corporate governance is worth a closer look. Shleifer
\& Vishny (1997) argue that legal protection of investor rights is
one essential element of corporate governance. Concentrated ownership,
through large share holdings, takeovers, and bank finance, is also
a nearly universal method of control that helps investors influence,
firm decision making. Although large investors can be effective in
solving the agency problem, they may also inefficiently redistribute
wealth from other investors to themselves.

Any talk about legal matters is incomplete without a mention of Bastiat\textquoteright s
(1968) timeless essay, The Law, which suggests the precise limits
under which the law has to operate; otherwise, the result would be
legalized plunder. The main purpose of law is to render justice, Sandel
(2010), illustrates the subjectivity and the subtleties that can arise
when dispensing justice.
\end{doublespace}
\begin{conjecture}
\begin{doublespace}
\textbf{Even if we take it for granted that the business of a business
is simply making more money, and skip the entire debate of whether
a firm (and hence the individuals comprising the firm) should identify
itself with a more purposeful aspiration, what we can see is that
with increasing complexity, it becomes harder to verify whether the
goals of the individuals are aligned towards the common objective.}\end{doublespace}

\end{conjecture}
\begin{doublespace}
Grant (1996) identifies the primary role of the firm as integrating
the specialist knowledge resident in individuals, into goods and services.
The task of coordination, which can be onerous and the possibility
of goal conflict are mentioned. The knowledge-based approach offers
a theoretical basis for understanding recent organizational trends,
including the development of new organizations forms which are more
horizontal, with fewer layers, greater empowerment, more team based
structures and inter-firm alliances. If the primary resource of the
firm is knowledge, if knowledge is owned and can only be exercised
by employees, the foundations of a shareholder value approach, with
distinct owner and operator, are challenged. This analysis fails to
account for is a more comprehensive approach embracing both knowledge
creation and application. Also, in larger settings, a specialist centric
view could lead to situations, where people successfully try and portray
a false image using titles, appearance, and other methods of superficial
perceptions reliant signaling (Kashyap 2010).

Carroll (1991) takes a starkly contrasting view and lists the many
social responsibilities of firms in their decreasing order of importance:
economic, legal, ethical and philanthropic. He mentions that for such
a reality to happen firm executives need to be moral instead of amoral
or immoral, which is again a hark back at the need for philosopher
kings. Friedman suggests that if shareholder value is maximized, then
the shareholders can heed to the call of non-profit generating responsibilities,
with their share of the profits. It is easy to see that the common
ground for both viewpoints is to have shareholders more involved with
firm decisions, which can happen better in smaller, employee owned
and operated firms.

Anderson \& Warkov (1961) summarize findings from previous studies
that assert that growing density of population in a society results
in increasingly complex forms of organization and that an increase
of size necessitates more complex forms of communication. Some studies
claim that, in addition to its effect on organizational complexity
growth also brings about a disproportionate increase in the size of
the administrative component. Finally, more and more complex tasks
may require that the coordination of an organization's differentiated
components be accomplished by an increasingly larger administration.
By studying the size, complexity and related characteristics of hospitals,
they find that: the relative size of the administrative component
decreases as the number of persons performing identical tasks in the
same place increases (in sharp contrast to other previous studies);
the relative size of the administrative component increases as the
number of places at which work is performed increases; the relative
size of the administrative component increases as the number of tasks
performed at the same place increases (or as roles become increasingly
specialized and differentiated).

Hall, Johnson \& Haas (1967) find that the relationships between size
and other structural components are inconsistent, similar to previous
research, which utilized size as a major variable. They review other
studies that consider the centrality of complexity within organization.
Large size is not in itself a critical characteristic of organizations.
Rather what appears to be important here is complexity, which is often
indicated by size but is quite distinct from it. One way of ascertaining
complexity is by measuring the number of occupational specialties
and the length of training required by each. The greater the number
of occupations and the longer the period of training required, the
more complex the organization. With increased size, the structure
of the organization becomes much more complex. The division of labor
becomes more differentiated and specialized; more levels of supervision
are introduced to maintain coordination and control; and more people
become involved in organizational planning. There are wide variations
in complexity and it is a structural condition which itself contains
a number of components.

Another definition of complexity, which appears to encompass the considerations
discussed above, is the degree of internal segmentation, that is,
the number of separate \textquotedbl{}parts\textquotedbl{} of the
organization as reflected by the division of labor, number of hierarchical
levels, and the spatial dispersion of the organization. There is a
slight tendency for larger organizations to be both more complex and
more formalized. More hierarchical levels are found in larger organizations
and relatively strong relationships exist between size and the formalization
of the authority structure. Their research, which included organizations
ranging in size from six members to over 9,000 members, representing
a wide range of types, such as educational, commercial, military,
governmental, manufacturing, religious, and penal organizations, suggests
that size may be rather irrelevant as a factor in determining organizational
structure.

Kasarda (1974) examines the structural implications of social system
size on three levels of the social system hierarchy: the institutional,
the communal, and the societal. He finds that size has a pervasive
influence on the internal organization of social systems. Size promotes
greater administrative intensity in institutions, communities, and
industrialized societies. As social systems expand, substantially
greater proportions of their personnel are devoted to communicative
or clerical functions. Another inference which may be drawn is that
large size promotes an increase in the proportion of professional
and technical specialists to handle the additional problems of information
gathering, evaluation and planning. Large size does reduce the proportion
of managers in an organization; but it raises the relative proportion
of other administrative personnel. The result is that the marginal
savings in management overhead are exceeded by the marginal costs
(in terms of man power) of larger clerical and professional staffs.

Rizzo, House \& Lirtzman (1970) find that measures of role ambiguity
and conflict within organizations can exist as two separate and independent
dimensions, with some correlation in expected directions with other
variables of importance to organizations.

Despite the number of interesting studies that have been carried out,
it seems to indicate that the results are inconclusive, which is not
surprising, given the complexities involved in the study of social
systems. These studies can be thought of adding to two viewpoints:
one is that with larger systems, complexity increases and a greater
proportion of people are necessary to co-ordinate and ensure that
the objectives are aligned; the other being that the administrative
component does not necessarily increase in proportion to size. The
main issue with both these contrasting viewpoints is that the span
of control or the sphere of influence of certain people will enlarge
and this mushrooming clout can be misused and even if it is not, inadvertent
mistakes can have staggering consequences.
\end{doublespace}
\begin{conjecture}
\begin{doublespace}
\textbf{It is worth considering whether the complexities of larger
systems might itself render such systems beyond the reach of proper
analysis and whether the social systems will evolve and change as
studies emerge suggesting a particular finding. }

\textup{That the uncertainties involved are manifold and they multiply
as systems grow is probably the only thing we can be certain about.
In the absence of a situation where all the individuals in a social
system are righteous, capable of divining the right information and
making perfect decisions, it becomes necessary to have smaller organizations,
with reduced scope in what they control, so that the effects of mistakes
or behavior deviating away from the goals, tends to be restricted.}\end{doublespace}

\end{conjecture}
\begin{doublespace}

\subsubsection{\textcolor{black}{High Energy Low IQ Syndrome}}
\end{doublespace}
\begin{conjecture}
\begin{doublespace}
\textbf{\textcolor{black}{The wider the set of responsibilities a
person faces, the lesser the attention each particular facet gets.}}\end{doublespace}

\end{conjecture}
\begin{doublespace}
\textcolor{black}{As an organization gets larger, it is reasonable
to assume that more issues will crop up that require resolution by
its most capable members. For simplification, here we can assume that
capability means depth of thinking that considers an array of possibilities
before a decision is made, something that a philosopher King would
be suited to do, as we alluded to earlier. We also emphasize that
IQ here does not necessarily reflect innate intelligence but is more
about the attention span one devotes to solving problems (hence the
amount of intelligent thought process). Few of us want to make wrong
decisions consciously. It is just situations and role models that
bring out the worst in us. Assuming then that the most capable members
are at the top levels, where the key decisions are made, we see that
the amount of thought each issue gets will be limited since many issues
are bubbling up to the surface. We then see that decision makers need
to develop an ability to shuttle between different demands, which
by definition is lower IQ. }

\textcolor{black}{This leads to the contradiction that the members
of the organization that persist in deeply analyzing issues and setting
the more appropriate direction forward do not reach the higher levels
since they are not displaying the energy or the capacity to multi
task among a myriad of concerns. The people that end up at the top
would be those that can make superficial, a.k.a quick and dirty, decisions.
Situations that are more prone to conflict are discussed later highlighting
that sub optimal decisions can be good enough in some cases, as long
as their impact is curtailed. Needless to say, this discussion overlooks
many things to convey the message that the margin for tolerance is
high for most day to day situations, where quick heuristic decisions
would be perfectly fine. The incidents that can differentiate good
and great decision making are far and few between. But if such situations
do arise, with the ever growing size of firms, it is likely that the
best decisions makers are not acting on it.}

There is the possibility that high energy decision makers rely on
high IQ lieutenants for advice. Even after assuming no loss of information
or dilution of directives across layers, no conflict of outcomes and
hidden agendas (all three are more likely in more complex conflicting
environments), various constraints related to chain of command and
final authority set in and if differing perspectives exist, it is
likely that high energy low IQ decisions might prevail.
\end{doublespace}

\begin{doublespace}

\subsection{The Need for Smaller Organizations}
\end{doublespace}

\begin{doublespace}
Diamond (1997) tracing the historical development of the trend towards
bigger organizations, supported by bigger communities, leads us to
the key stimulus that was the surplus generated by the superior modes
of agricultural production. This made possible the establishment of
a non-producing class, whose members were crucial for the rapid development
of writing, science, cities, technology-based military prowess and
formation of states. Dense population centers that could be supported
near these lush agricultural centers had a greater exchange of ideas,
bringing new innovations into force and also allowing the extraction
of rents from a larger number of individuals who came to depend on
these new products that were fed by the invention spree. 

The blessings of large population centers, on the sciences and the
arts, have been tremendous. Development of regions like Silicon Valley
in California or Broadway in New York is due to the rapid exchange
of ideas. While the benefits of dense populations accrue up to a certain
point, the negatives of overcrowding, shortage of resources and diminishing
returns set in after a certain stage, giving rise to increasing disparities
between the residents in these packed colonies. The widespread use
of technology to connect people facilitates interaction among relatively
far flung dwellings, removing the need for the congregation of individuals
to accelerate the pace of evolution of human civilization.

Damanpour (1996) considers the relationship between innovation and
two major indicators of organizational complexity-structural complexity
and organizational size. The study finds that the association between
structural complexity and innovation depends upon operational definition
of complexity, environmental uncertainty, use of manufacturing organizations,
use of service organizations, focus on technical innovations, focus
on product innovations, and focus on implementation of innovation;
and the association between organizational size and innovation depends
upon operational definition of size, environmental uncertainty, use
of service organizations, use of for-profit organizations, focus on
technical innovations, and focus on product innovations.

Busenitz \& Barney (1997) examine the differences between entrepreneurs
and managers in large organizations with respect to two biases and
heuristics: overconfidence (overestimating the probability of being
right) and representativeness (the tendency to overgeneralize from
a few characteristics or observations). Entrepreneurs and managers
think differently, behave differently and utilize heuristics and basis
to different degrees. If the use of biases and heuristics are stable
over time, then those who are uncomfortable with heuristic based decision-making,
on more occasions, will be attracted and selected into larger organizations
and vice versa.

Under conditions of environmental uncertainty and complexity, biases
and heuristics can be an effective and efficient guide to decision-making.
The counter intuitive message here is that larger organizations can
give rise to complexity and might in fact require a heuristic approach
and be ill-equipped to handle circumstances that require decision
making shortcuts, due to the more status-quo loving or risk averse
nature of its managers. Whereas the use of cognitive biases may be
beneficial and required in some circumstances, it can lead to severe
and systematic errors in others. 

The implication of this is that large organizations are generally
more stable than smaller ones; but the price of failure is greater
and hence they are suitable, or perhaps, are more tolerable, only
for stable environments and certain areas of society. If entrepreneurial
attitudes and risk taking abound in organizations that can negatively
impact the lives of people, it is better to ensure that they are contained
in the harm they can do. 
\end{doublespace}
\begin{conjecture}
\begin{doublespace}
\textbf{Highlighting the greater damage that can result when things
go wrong with large systems might seem like a negative threat based
approach. The rationale here is that for normal functioning, either
small or large can be effective and adequate enough; but when the
unexpected happens, the results are more severe and widespread with
larger systems.}\end{doublespace}

\end{conjecture}
\begin{doublespace}
Starting with the premise that profit maximization is the driving
force of a business, means accepting that a business requires the
division of a finite amount of wealth among its participants, a situation
inherent with possibilities for conflict. This suggests that it would
be prudent to restrict the scope of organizations that require allocation
of resources, for which agents do not necessarily have limited desirability
(alternately stated as, diminishing marginal utilities won't set in
till a very high threshold is reached). Comparing large universities
and large businesses should provide more clarity. A business needs
to create wealth and share that among its employees. A university
needs to create knowledge, which is easier to share than wealth. This
leads to the conclusion that universities can be big without giving
rise to undue conflicts while businesses need to be smaller.

Porta, Rafael La, et al (1996) argue that trust is an important ingredient
to ensure co-operative behavior. They further mention that while trust
might be easier to establish in smaller settings where repeated games
can be played that present opportunities to seek corrective actions
for previous wrongs, even in larger societies with a greater level
of trust, co-operative outcomes can be observed. What this tells us
that all else being equal, a smaller setting is better for coordinating
efforts.

Smaller size means that it would be easier to check, what, is being
done by the institutions and the people involved. We need to look
at the argument that if organizations are small, there would be many
such institutions and hence many of them to monitor, making it a harder
task. But given the reduced scope of smaller organizations, there
would be greater transparency, less complexity and a stronger relationship
between the service provider and the served.
\end{doublespace}
\begin{conjecture}
\begin{doublespace}
\textbf{The smaller size leads to more number of superior quality
interactions between the same parties, leading to a repeated game
setting, which is known to produce more co-operative behavior. The
strengthened relationship effectively acts as an enforcement agent
towards both the parties.} \end{doublespace}

\end{conjecture}
\begin{doublespace}
The smaller institution cannot extract large rents for itself, since
otherwise it would cease to be competitive against the myriad number
of smaller institutions that are available as alternatives and such
actions would deter people from doing business with it. The people
that are benefiting from the services of an institution would be under
close scrutiny from the institution, which bears the burden to ensure
that its services are put to the best use possible, since that is
integral for its own prosperity. Agents are driven against myopic
self-motivated behavior, since maximal benefits accrue by acting with
a longer-term vision. 
\end{doublespace}
\begin{conjecture}
\begin{doublespace}
\textbf{Any set up where the players involved have a fundamental incentive
to be on best behavior, functions better than other alternate possibilities.
Smaller size reduces complexity in many ways and makes it harder to
hide things under the rug.}\end{doublespace}

\end{conjecture}
\begin{doublespace}
This also makes it harder for corruption or other illegal episodes
to happen. Systemic failures, wherein most organizations in a sector,
are severely affected in a negative way, are less likely, since we
have many small organizations and the degree of interconnectedness
will be lower that in a set up with a fewer number of organizations.

As organizations grow bigger, a greater share of the individuals that
are part of it, become involved in just making sure that the organizations
are running smoothly. This takes people away from becoming involved
with the actual generation of ideas or producing a tangible output
or adding to real growth and welfare. Instead of excessive resource
allocation to ensure co-ordination and control, we simply need more
transparency, which will result in more fairness and the right thing
being done. Honesty is not entirely innate; it can be instilled and
it follows from the recognition that human conduct is usually a response
to the incentives and the situations. While, formal attempts at tracing
the impact of integrity on the functioning of large institutions are
worthwhile; a simpler argument, that smaller organizations with less
complexity create a better alignment of incentives and situations,
and give rise to an environment where it is harder to hide immoral
incidents, and foster more righteous behavior, can be shown to hold
water.

We could raise the point that the compensation of executives in large
organizations can be monitored and going this route would be easier
than having to monitor thousands of smaller firms. The rebuttal for
this would be that when someone has access to large amounts of money,
the chances of misappropriation are higher than when there is no access
to large amounts. The recent financial crisis had instances where
large bonuses were paid out even by firms that were receiving bail
out funds from the government, under the excuse of retaining talent,
among others. 

While it is not entirely inappropriate to impose limits on executive
compensations, it is highly likely that clever ways to derive excessive
compensation will be devised when there is possibility of being able
to siphon large amounts of money. This also raises the question of
the system of governance. Simply put, this is about whether the state
should interfere with the specific of how a firm is run or should
the state restrict the main activities of a firm. The next section
also considers this in further detail, but without deviating much
further from our discussion, we can say that giving the state power
over every day affairs can be disastrous.
\end{doublespace}
\begin{conjecture}
\begin{doublespace}
\textbf{Does size matter? }Does large size lead to stability? Turning
to nature again for inspiration, we don\textquoteright t see excessively
large organisms, despite some creatures that never stop growing. Similarly,
organizations have a tendency to grow. We are a growth obsessed society.
A mindset that tolerates the omnipresent stressor of competition and
celebrates the birth and death of organizations, helps prevent abnormal
growth. The pseudo-stability of big organizations can cause disasters
when they fail, since most systems can cope better with many small
continuous demises than a few large sudden deaths.\textbf{ Size does
matter.}\end{doublespace}

\end{conjecture}
\begin{doublespace}
There is also the possibility of bigger organizations hiding inefficient
parts within themselves by subsidizing their existence. The argument
about economies of scale is not as applicable today, because, we use
automation and machines to a great extent for agriculture and producing
goods. Organizations are knowledge based as opposed to the traditional
manufacturing industry, for which such a production term needs to
be applied. If organizations are to be small, it is helpful to have
a climate that facilitates entrepreneurial activity and allows the
easy birth and growth of new businesses. 

Acs and Varga (2004) highlight two important proxy measures of the
existence of entrepreneurial opportunity, the tendency of people to
engage in self- employment and the tendency of people to start new
firms. Using data from the Global Entrepreneurship Monitor (GEM) project
they examine the relationship between entrepreneurship, knowledge
spillovers and economic growth. There are manifold ways to measure
entrepreneurial activity. One overbearing dissimilitude is between
opportunity-based entrepreneurial activity and necessity-based entrepreneurial
activity. Opportunity entrepreneurship represents the voluntary nature
of participation and necessity reflecting the individual\textquoteright s
perception that such actions presented the best option available for
employment, but not necessarily the preferred option. Opportunity
entrepreneurship differs from necessity by sector of industry and
with respect to growth aspirations. Opportunity entrepreneurs expect
their ventures to produce more high growth firms and provide more
new jobs. Any gauge of entrepreneurial activity needs to factor in
this distinction.

Bockstette, Chanda and Putterman (2002) construct an index that captures
the length of state experience. This index is higher for countries
that have a longer experience with state-level institutions and such
countries have higher political stability, institutional quality and
economic growth. It would be advisable to supplement the level of
entrepreneurial activity with a measure that captures the depth and
history of entrepreneurial culture present in a country. 

Beck \& Demirguc-Kunt (2006) find that small firms face larger growth
constraints and have less access to formal sources of external finance,
potentially explaining the lack of their contribution to growth. This
highlights the issue that larger organizations could crowd out smaller
ones, since the pseudo-stability they display, will guzzle away resources
from smaller organizations. 

Kashyap (2015a) looks at the financial services sector and ways in
which the increasing size of financial systems leads to greater profits
for the sector at the cost of stability and even perhaps a deviation
from the core functions of the sector. While there are numerous ways
to monitor and overcome this departure from the intended objectives,
a self-reinforcing way that reduces complexity is decreasing the size
of financial firms.

Linck, Netter, \& Yang (2009) argue that the requirements of the Sarbanes
Oxley act have increased the demand and reduce the supply of directors,
and there is a potential adverse impact on smaller public firms due
to increased compensation burden. The increased monitoring burden
imposed upon organizations is working to the detriment of smaller
organizations, which might need lesser surveillance than their larger
counterparts.

Acs and Varga (2002) hypothesize that any spatialized theory of technology
led regional economic growth needs to reflect three fundamental issues.
First, it should provide an explanation of why knowledge related economic
activities start concentrating in certain regions leaving others relatively
underdeveloped; second, it needs to answer the questions of how technological
advances occur and what the key processes and institutions involved
are; and third, it has to present an analytical framework in which
the role of technological change in regional economic growth is clearly
explained.

Beck (2008) mentions that bank size is positively correlated with
complexity so that large banks are harder to monitor than small banks.
De-Nicolo (2000) argues that bank consolidation is likely to result
in an average increase in banks insolvency risk.

This prong is the most important of the four since if we get this
right, the reduced size and complexity of any organization or system;
helps realize the limitations and aids in the detection of any deviations
from the expectations; ensures that the responsibilities and incentives
of the parties involved are aligned and continue to stay aligned with
the original targets; and makes it easier to ascertain the unintended
consequences of any efforts, which are hard to completely eliminate,
as we will see in the next section.
\end{doublespace}

\begin{doublespace}

\subsubsection{Organizations as an Amalgamation of Lego Blocks}
\end{doublespace}
\begin{conjecture}
\begin{doublespace}
\textbf{We are building increasingly complex organizations and we
are using increasingly complex methodologies to study them. The complex
methodologies are just rearrangements of simpler rules just as complex
organizations are aggregations of simpler blocks. }

\textup{\textcolor{black}{This suggests that the smaller the bunch
of Lego blocks (Endnote \ref{enu:Lego Blocks}) used, the easier will
it be to understand any structure that results from combining them.}}\end{doublespace}

\end{conjecture}
\begin{doublespace}

\subsection{Redirecting Unintended Outcomes}
\end{doublespace}

\begin{doublespace}
Paiche and Sterman (1992) inquire into decision making in complex
environments and conduct an experiment where subjects must manage
a new product from launch through maturity, and make pricing and capacity
decisions. Building upon previous studies, they demonstrate that decision
making in complex dynamic environments tends to be flawed in specific
ways by not accounting sufficiently for feedback loops, time delays
and nonlinearities. Even with a decent amount of experience, there
is no evidence that environments with high feedback complexity can
produce improved decision making ability. Poor decision making in
complex production systems can create pervasive booms and busts, where
new products can have exponential sales increases, fueling rapid growth,
often leading to overcapacity, price wars, and bankruptcy.

Pollay (1986) reviews theories about advertising\textquoteright s
social and cultural consequences. Advertising is viewed as intrusive
and environmental and its effects as inescapable and profound. Advertising
is seen as reinforcing materialism, cynicism, irrationality, selfishness,
anxiety, social competitiveness, sexual preoccupation, powerlessness,
and/or a loss of self-respect. 

Kashyap (2015b, 2015c) looks at unintended consequences as it applies
to the financial services industry and the difficulties in being able
to anticipate the outcomes of complex systems. Norton (2002) mentions
the law of unintended consequences, as often cited but rarely defined,
as actions of people -and especially of government- always have effects
that are unanticipated or unintended. Building on the work of others,
primarily, Merton (1936), the discussion labels various sources of
such unintended consequences: ignorance, error, instances where individuals
want the intended consequences so badly they choose to ignore any
unintended effects and self-defeating predictions where the public
prediction of a social development proves false precisely because
the prediction changes the course of history. Government attempts
at reform have the largest scope and hence perhaps the actions of
the government and politicians are alluded to directly in the article.

Ash, Berg \& Coiera (2004) outline a number of issues within a framework
describing two major kinds of silent errors caused by health care
information systems: those related to entering and retrieving information
and those related to communication and coordination. The potential
causes of these errors are subtle but insidious. They argue that many
of these errors are the result of highly specific failures in patient
care information system design and/or implementation. They do not
focus on errors that are the result of faulty programming or other
technical dysfunctions. Hardware problems and software bugs are more
common than they should be, especially in a high-risk field such as
medicine. 

However, these problems are well known and can theoretically be dealt
with thorough testing before implementation. Similarly, they do not
discuss errors that are the result of obvious individual or organizational
dysfunction such as a physician refusing to seek information in the
computer system, \textquoteleft because that is not his task, or a
health care delivery organization cutting training programs for a
new information system for budgetary reasons. They focus on those
often latent or silent errors that are the result of a mismatch between
the functioning of the information system and the real-life demands
of health care work. Such errors are not easily found by a technical
analysis of the system design, or even suspected after the first encounter
with the system in use. They can only emerge when the technical system
is embedded into a working organization and can vary from one organization
to the next.

Schoorman (1988) examines the tendency of supervisors to escalate
their commitment of a previously expressed opinion by biasing performance
ratings in the context of a real organization. There was strong support
in the study for two hypotheses: (a) that supervisors who participate
in a hiring or promotion decision and agree with the eventual decision
would positively bias subsequent performance appraisal ratings for
that employee; and (b) that supervisors who participate in the original
decision but disagree with the decision would bias subsequent performance
appraisal ratings in a negative direction.

Cornelius (2001) assesses the efficacy of the strategy of immigration
control implemented by the US government since 1993 in reducing illegal
entry attempts, and documents some of the unintended consequences
of this strategy, especially a sharp increase in mortality among unauthorized
migrants along certain segments of the Mexico-US border. The available
data suggest that the current strategy of border enforcement has resulted
in re-channeling flows of unauthorized migrants to more hazardous
areas, raising fees charged by people-smugglers, and discouraging
unauthorized migrants already in the US from returning to their places
of origin. However, there is no evidence that the strategy is deterring
or preventing significant numbers of new illegal entries, particularly
given the absence of a serious effort to curtail employment of unauthorized
migrants through work-site enforcement.

The Sweeney and Sweeney (1977) anecdote about the Capitol Hill baby-sitting
crisis exposits the mechanics of inflation, setting interest rates
and monetary policies required to police the optimum amount of money.
The creation of a monetary crisis in a small simple environment of
good hearted people expounds that even with near ideal conditions,
things can become messy; then in a large labyrinthine atmosphere,
disaster could be brewing without getting noticed and can strike without
much premonition. This emphasizes the need to keep complexity at bay
and establishing an ambience where repeated games can be played with
public transparency, so that guileful practices can be curtailed.

All of this tells us that intended outcomes could churn out negative
side effects and the reverse could happen as well. But to create a
robust society we need to minimize the impact of the downfall. The
topic discussed earlier about having better metrics to measure welfare
would be helpful to identity unintended outcomes.

If governments or any organizations start intervening to set strict
limits on production, on the amount and manner in which people consume
and live, the results could be catastrophic and lead to too much state
control or to socialism or even communism. (Hayek 2009; Marx, Engels
1959; are seminal works). The other extreme to this is completely
free markets, or capitalism, which as we are realizing will lead to
huge inequalities in society. While it is hard to draw a strict line
between these two modes of governance, we need elements from both
models of governance and economic policy. Our earlier discussion on
the advantages of smaller size firms applies here, wherein, with the
prevalence of smaller organizations, incentives would be aligned such
that it becomes easier to spot wrongful conduct, leading to better
governance with lesser monitoring.
\end{doublespace}

\begin{doublespace}

\section{Discussion}
\end{doublespace}

\begin{doublespace}

\subsection{Small Business as a Solution to Corruption.}
\end{doublespace}

\begin{doublespace}
When power and wealth are concentrated in the hands of a few, it usually
leads to corruption and exploitation of the masses by a few. This
is inevitable and follows from human nature and the seductive allure
of power. While taming human nature is possible, let us leave that
out for now since it is a somewhat arduous task that could span generations.

Every attempt at social reform has been to address this issue. This
exploitation of the masses by a few happens, in the extreme case,
even with both the two leading models of society and governance namely,
Socialism and Capitalism. A more practical and immediate solution
lies in ensuring that the basic needs of any individual are met and
he/she has a safety net that ensures a certain minimal standard of
living. When survival becomes easy, people will not give in easily
to exploitation. This can happen through a grassroots movement by
ensuring that small businesses and mid-size businesses thrive and
the barriers to entry for such new organizations in an economy are
minimal to non-existent.

When many such small organizations succeed and new ones can crop up
easily, power tends to get less concentrated and therein lies the
true solution to the many ills faced by any society. Having said this,
we do not mean that we should not say anything against corruption
or other ills. In fact, any effort by any individual against corruption
or the like should be commended and any such effort is certainly not
in vain. But what we really need to understand is that we need to
support organizations that help small businesses grow and thrive.
This is what the Great Mahatma Gandhi also stressed in his vision
of every village meeting its needs indigenously through small scale
industry. This can apply to all countries around the globe even though
there are varying levels of corruption and other cultural nuances.

A simple analogy is when a person is very sick, medicines help, but
what is needed longer term are better diet, exercise and living conditions.
Same goes with creating more rules and protests, they are helpful
temporarily but what is needed longer term is an environment where
individuals can establish self-sufficiency easily thereby reducing
the chance for them to get exploited. When this happens, the common
man is less dependent on the Government and other large organizations,
such organizations will have less power making it more difficult to
be corrupt or abusive.

Surely, this perfect recipe for a smooth functioning society is easier
said than done. Most would easily see that the simplification in this
narrative is for the brevity that is essential in any call for action,
which in this case, is for individuals and organizations to realize
the need to promote small and medium enterprises, their responsibility
in making this happen and the benefits that will accrue to them once
this starts happening. As with any daunting challenge, there will
be opposition from certain incumbents that stand to lose the most
with the added risk that success might seem completely elusive or
become available in small unsatisfying doses for a long time, there
is simply no excuse for not trying.

While this might seem like an attempt to change the world, it is sheer
arrogance, possibly bordering stupidity, to think we can change the
world that has existed for millions if not billions of years in our
blink of a lifetime. The simple belief espoused here is that promoting
small-biz is one way to make life in a society a fun ride for a lot
of people. The other option might, of course, be to resurrect the
psychedelic 60's, the hippie lifestyle and the blissful ignorance
of LSD, all of which is neither recommended nor disapproved.
\end{doublespace}

\begin{doublespace}

\subsection{The Existential Question!!! }
\end{doublespace}

\begin{doublespace}
\textbf{\textit{Why do we need organizations that aid the development
of smaller organizations and businesses?}} The one sentence that can
answer this question is, 
\end{doublespace}
\begin{conjecture}
\begin{doublespace}
\textbf{\textquotedbl{}Necessity is the mother of all creation, but
the father that is often forgotten is frustration\dots \textquotedbl{}}\end{doublespace}

\end{conjecture}
\begin{doublespace}
It is necessary for organizations, which are seeking to help small
businesses flourish, to exist. The earlier section on \textquoteleft Small-Biz
as a Solution to Corruption\textquoteright{} has a more detailed explanation
of this necessity. While that is the necessary part, the frustration
that should bring about such organizations is the increasing explosion
of information, jargon and the deception of society by the so-called
experts. We mean no disrespect to any experts who have spent many
years working in a field. We definitely include people doing Research
and Development in the expert category. We as a society, clearly need
both. There is also much to be said about the instincts that people
tend to develop when they engage in a certain behavior for long periods
of time. But, what we need to realize is that in most situations that
do not warrant a highly specialized outlook, we value expert opinions
a lot more, even though they may not really be all that better than
the opinions of an average person.

It is common to fear darkness and look towards the light and hence
the nexus to Diwali, the festival of lights, as illuminated by the
following metaphor. As we celebrate the festival of lights, it is
worth pondering about one interpretation of light, which stands for
knowledge that dispels darkness, which is representative of both ignorance
and over confidence.

Needless to say, we need to mitigate this two-sided problem, where
on one side we know nothing of too many things and on the other side,
we feel we know everything about some other things, by spreading knowledge
about varied concepts and by demonstrating the limitations of whatever
we know.

Summing it up, we need organizations to promote small businesses and
continuously educate people. This relates to the earlier discussion
of the central functions of organizations, which is the creation and
application of knowledge to solve the problems facing society. While
it is hard to draw a strict boundary between the two streams, gaining
knowledge can represent efforts at creating and disseminating interesting
thoughts, and problem solving could be the work done with small businesses
in resolving their concerns.

In Conclusion, \textbf{\textit{Organizations like the one above need
to exist, in this information age, to try to uncover nuggets of knowledge
amidst buckets of B.S (Endnote \ref{enu:B.S.-stands-for}).}}
\end{doublespace}

\begin{doublespace}

\subsection{A Way of Life}
\end{doublespace}

\begin{doublespace}
Why stop at the organization level, when a culture that is constantly
involved in gaining knowledge and applying this knowledge to solving
problems can be a beneficial lifestyle. Gaining knowledge and solving
problems need not be a trade, but it can simply be something people
enjoy doing; leading to the motto, \textbf{\textit{Not just a business,
but a way of life}}.

What other values would be compatible with such a tradition? Such
a heritage would challenge conventional thinking by claiming that
most complexity is the result of viewing the world from a highly specialized
point of view and the increased use of unnecessary jargon. It would
continuously strive to find common elements among various seemingly
unrelated disciplines. This would be done, while being fully aware
that all situations have certain unique elements to them and that
recognizing this is the key to arriving at a more comprehensive resolution
of problems. Years of experience are not useless. Previous experience
can be extremely valuable, but only when used where it is really needed.
Respect for every bit of knowledge gained from every possible source
is important; what is more important is an attitude that does not
take this knowledge too seriously. This would mean gaining knowledge,
so we can forget all about it when we have to and recollecting knowledge
from a myriad of sources when we want to. If this sounds confusing,
congratulations, this is the right track; confusion is the beginning
of understanding.

This lifestyle would seek to help all kinds of organizations by advocating
a somewhat unique approach to problem solving. When organizations
realize they are facing some issues, they look for people either internally
or externally who can solve it.
\end{doublespace}
\begin{conjecture}
\begin{doublespace}
\textbf{A lot of times, the main issue about the issue is that it
is not clear what the real issue is.} \end{doublespace}

\end{conjecture}
\begin{doublespace}
The reason for this is very simple, when an issue is identified, it
is either done by someone who does not have the time to solve it or
does not know how to solve it or simply does not want to deal with
it themselves. This could very well be due to the many valid constraints
they face in their current organization. So they make their best effort
to isolate the issue. Then, what is usually done is the issue is bucketed
into a few standard categories. Then an expert in that category is
brought in to take care of it. The expert here is someone who has
dealt with similar issues many times before and has experience dealing
with such problems for a number of years.
\end{doublespace}
\begin{conjecture}
\begin{doublespace}
\textbf{So at this point, we are not sure what the issue is and we
have an expert on this issue looking at it. }\end{doublespace}

\end{conjecture}
\begin{doublespace}
When we talk about experience over a number of years in the conventional
sense, it does lead to expertise and specialization. This means when
a specialist is looking at a problem, he/she is looking at it in one
way, identical to an approach he might have used before, and might
provide a solution quite similar to previous situations that might
have some common elements to the current problem. This happens in
a lot of organizations, a lot of times. The specialist has the advantage
of being the most knowledgeable person in the initial diagnosis phase
and hence someone who is not a specialist does not seem to add much
value in the initial phase and the opinions of few such non-experts
get serious consideration in this initial phase. We might have seen
this play out many times before in almost any organization we have
been part of. The good part is that this problem can be solved with
a certain way of problem solving.

Dealing with problems differently is an attitude, a frame of mind.
We mean no disrespect to people who are passionate about something
and spend years pursuing it. But, the nature of the issue we are facing
is such that, in many situations that do not require a lot of expertise,
we value the opinions of experts, even though it might not contribute
anything materially significant. We can have a bright high school
student, talk as intelligently about the economy, as an economist,
who has been doing this for two decades. The same goes with say, an
expert in marketing, accounting, law or a number of other disciplines.
(While it is tempting to generalize this and say that with the right
amount of training, this applies to almost every discipline; we steer
away from doing that here.) There is also an element here about how
luck is important and this can be a long discussion and we will leave
it out for now.

The takeaway from this is that people overestimate the amount of additional
knowledge that expert\textquoteright s possess. The bias that forms
naturally, when experience is used to solve problems, gets overlooked.
Having said that: when looking at a problem, it is not just helpful
but absolutely essential, to not muddy the issue too much with our
previous experience. This does not mean that we have to leave out
experts from our teams. It is certainly important to draw lessons
from previous situations, to work with many experts and solicit their
advice. But in the initial phase, we should try to get as many contrasting
opinions out on the table as possible. We should also try to find
common elements between a particular problem, and say some other problems
the organization might be facing in other related divisions or groups.
Easy as this might sound, to fully get a team working like this, takes
training and a great deal of modesty from the members of the team.
To be able to accomplish this requires the right approach to problem
solving and continuous honing of those skills. This way of Problem
Solving provides better satisfaction for all the stakeholders.

It is worthwhile to mention the kind of persons to keep company with,
to promote this lifestyle. It would be someone who knows not to take
their experience as their greatest asset. The first thing to learn
is to be able to drop our baggage of experience at the door step before
looking at any problem. A fresh perspective and untainted observation
should form the basis of getting trained to diagnose any situation.
Next, it is about learning to apply the knowledge gained from sources
as obscure as watching a clown perform, in a busy Paris street or
a day spent on an African Safari to the problem at hand. It is a constant
involvement in gaining knowledge and questioning what we learn. This
critical evaluation of knowledge helps to find patters that are otherwise
not easily noticed. It is more than surprising, when one discovers
how things are interrelated and the subsequent \textquotedbl{}aha\textquotedbl{}
moment that comes about when we uncover these relationships. It is
important to be extremely organized in our thoughts and efforts; but
given what has been discussed, there is no standard life cycle or
project plan that will apply and we can follow. We need to observe
and take detailed notes in this first phase of analysis. This makes
the initial phase of any project the hardest phase.

Another example of this thinking paradigm would be, people in the
operations or the inventory management department should not just
be involved but they should understand all aspects of developing new
products in the product development group. The common complaint to
this is that with only 24 hours in a day, and less than half of that
devoted to work, someone in operations does not have the time to learn
about new products. There is a simple and effective solution to this.
When new projects are started, in the initial phase, it becomes necessary
to demonstrate how cross departmental understanding can be achieved
with very minimal time consumption when the key stake holders are
incentivized to understand the importance and made to participate
in such efforts.

This leads to the most important aspect of how to tackle new issues
- Teamwork. The central belief for team work to work is: what one
person can do, two or more people can do better. Teams need to be
structured around this belief. Individuals that have dealt with problems
similar to the ones being faced need to be teamed with people with
expertise in other areas. The instincts of the expert coupled with
the fresh perspective of the so called non-expert will ensure that
no stone is left unturned in looking for solutions. Being effective
problem solvers means taking on different roles from being a sounding
board, to acting as coaches, to performing complete hands on implementations,
as the particular case may require. What is important is a desire
to get the most things done. This approach might pleasantly surprise
everyone involved, at how much more can be accomplished than was initially
thought possible. The result is what we can term \textquotedblleft A
Fundamental \& Original Approach to Problem Solving\textquotedblright .

Tying the discussion back to the central tenet of the paper and the
arguments put forth earlier, we can surmise that the way of life discussed
here is easier to follow when organizations are smaller. Corruption
will be less and objectives of the players are aligned with the intended
outcomes.
\end{doublespace}

\begin{doublespace}

\section{The Path Ahead}
\end{doublespace}

\begin{doublespace}
The core theme we have proposed is that organizations need to be small.
This opens up a number of questions, to which we provide partial answers
here, and point out that much further thought and work is required
to more satisfactorily assuage these objections.

How small is too small or how big is still too big? While on the surface,
not knowing the answer to this question might seem like a show stopper
for tangible action. As alluded to, in the introduction; a relative
scale in terms of what size works for what purpose, is what we can
hope to accomplish. The four pronged analysis technique can help reveal
where size is hampering the objectives of any particular social system.

What are some ways to ensure organizations become small and stay small?
How can we transform society in its current state with large organizations
to a state with smaller organizations painlessly or with minimum negative
consequences? How do we suppress the natural urge and desire we (including
owners and managers of organizations) have to grow our organizations?
What are the implications of smaller size for huge multinational organizations
and the area of international trade and finance? Is there really a
stability that comes from size or will the birth and death of organizations
of smaller organizations be accepted such that there will be a realization
that larger size only gives pseudo stability? This brings up the subject
of regulation and the limitations of regulation, not to mention, the
unintended consequences that always creep in along with the intended
ones. While not attempting to have some regulation in place would
be unwise, any attempt at regulatory change is best exemplified by
the story of Sergey Bubka\textsuperscript{}(Endnote \ref{enu:Sergey-Bubka}),
the Russian pole vault jumper, who broke the world record 35 times.
Attempts at regulatory change can be compared to taking the bar higher.
Despite all the uncertainty, we can be certain of one thing, that
people will find some way over the intended consequences, prompting
another round of rule revisions, or raising the bar, if you will.
What would work better in the long run is a cultural mindset that
takes smaller size as a blessing and worships the entrepreneurial
spirit that can continuously churn out smaller organizations.

The excellent reference by Creswell, J. W. (2013) has pointers on
how some of the above discussion can be shaped further using a mixed
methods approach, that is a mixture of both quantitative metrics,
supplemented with qualitative insights.
\end{doublespace}

\begin{doublespace}

\subsection{\textcolor{black}{The Compensation Ratio Test}}
\end{doublespace}
\begin{conjecture}
\begin{doublespace}
\textbf{\textcolor{black}{The ratio of the highest and lowest compensation
levels in an organization should not be greater than a certain factor. }}\end{doublespace}

\end{conjecture}
\begin{doublespace}
\textcolor{black}{As organizations age, the ratio should get smaller
which incentivizes newer organizations to start up. The ratio factor
though seemingly ambiguous can be related to the amount of wealth
at the disposal of the organization or the community it is part of
and the corresponding population (}Kashyap 2016b)\textcolor{black}{.
Such an approach could potentially lead to an extreme situation where
only a small percent of the population will be employed and controls
most of the resources. Another potential issue is that a temporary
workforce could be used and profits could be deferred after the work
force has been trimmed. This again highlights the issue that clever
workarounds are always possible, so keeping organization small and
simple should be the real objective. }

\textcolor{black}{Tax breaks for smaller firms, breaking up big companies
on antitrust grounds, greater blocks on merger activity and other
manners of enforcement can and should be employed but they will be
less effective since they are external to the organization. The compensation
ratio is an internal factor, constantly at work monitoring the proceedings,
resulting in a greater chance of success.}
\end{doublespace}

\begin{doublespace}

\subsection{Final Thoughts}
\end{doublespace}

\begin{doublespace}
If organizations end up becoming small, will this affect the progress
we have seen in society due to the various contributions from the
arrangement and synchronization of resources and efforts on a massive
scale? Businesses are needed to channel resources efficiently and
ensure that freeloading is minimized. While conceding that the maxim
of profit maximization is a necessary evil for now, the great strides
that civilization has taken, lends hope that perhaps some day agents
will not require a driving force to ensure that they reach their productive
best and perhaps passion alone can propel us forward. Till we reach
such a Utopian reality, we need businesses and we need to make sure
they are small.

An extreme situation that needs evaluation is whether we would be
a deprived society without the large scale alignment of efforts and
production. An example here, inspired by nature, is that of honey
bees (Winston 1991; Seeley 2009). To survive and get through the harsh
winter, requires co-ordination from a huge number of worker bees and
a tight knit community to collect nectar from far flung flower sources.
A handful of bees would perhaps not function as effectively, as a
typical hive that can have 20,000 all the way up to 200,000 individual
bees, in protecting their homes and gathering sufficient food. To
facilitate this high degree of co-ordination among many individuals,
bees have specialized functions determined right at the biological
level complemented with a sophisticated communication mechanism. 

Bhargava (1989) and Bühler (1886) discuss historical precedents where
society was grouped into divisions and specific functions were adopted
by individuals in each group. Huxley (2007) narrates an illustrative
story of a utopia in a futuristic fictional setting, where professions
are determined right at birth. These examples seem to highlight that
unless there are rigid restrictions on roles and responsibilities,
which are enforced either biologically or made culturally prevalent,
massive co-ordination might be harder to accomplish. Modern technology
has provided us with ample methods of effective communication and
with better means of communication; the optimal size might also increase.

Doidge (2007) presents classic cases from the frontiers of neuroscience
that chronicle the biological changes happening in the brain driven
by external impetuses, revealing that adapting to new circumstances
and learning to deal with adversity are almost hard wired into us.
In essence what they reveal is that the brain constantly changes as
situations change. (McDonald and Tang 2014) look at research from
the interdisciplinary field of social cognitive neuroscience that
provides insights as to how managers learn and develop, resulting
in theoretical propositions and practical implications. Neuroscience
offers potential to theoretically advance our understanding of management
development as well as practically enhance managerial capacity to
(a) reflect with a deeper sense of self-awareness, (b) analyze with
greater balance across hard and soft data, (c) position organizations
within broader perspectives, (d) collaborate inter-personally by establishing
relationships that engender egalitarianism and trust, and (e) enact
change in a nonlinear manner. This avenue of research holds a lot
of promise.

Surely it is a big ask that we solve all of society\textquoteright s
problems completely with one magic bullet. What we can realistically
hope as a starting point is at least a consensus as to which direction
society should move towards, for there to be better living conditions,
which we can argue is the main objective of society. If we are conceptually
in agreement that small is the sensible way forward, despite what
the answers to the other questions may be, we know in which direction
we should proceed. This means putting aside for the moment, the wonderful
advice of (Daft 1985) of mixing a theoretical approach with empirical
aspects in one paper to make it convincing for reviewers and editors,
in favor of a purely conceptual paper since; if we know what needs
to be done, how to do it becomes secondary since the many different
ways to accomplish something will lead to the same final outcome.
Any situation that requires co-ordinated efforts to shape and share
limited resources can produce conflict and in such situations complexity
needs to be reduced and the scope of conflict needs to be minimized,
by having smaller organizations.

Given the breadth of the hypothesis, mistakes are hard to avoid, much
further research and numerous revisions will need to be made to answer,
augment, modify and amplify these questions.
\end{doublespace}

\begin{doublespace}

\section{Conclusion}
\end{doublespace}

\begin{doublespace}
We looked at several examples that have intuitively substantiated
the coherence of the four pronged approach. 

By considering each prong in isolation and then finally integrating
the findings, we hope to establish different aspects of what would
be crucial to increasing welfare and also what would be the limitations
of any such recommendation. By recognizing the gap that exists between
the fundamentals that drive the behavior of individuals or institutions,
and the expected outcomes from their actions, we hope to highlight
how it becomes relatively straight forward to set incentives that
can maximize welfare.

If we start by reducing the size of institutions, it becomes easier
to monitor them and a certain level of self-governance is also put
into place. The other prongs then follow somewhat naturally and where
there are deviations from what is desired, the reduced complexity,
allows corrective mechanisms to be administered with less effort.
Highlighting the greater damage that can result when things go wrong
with large systems might seem like a negative threat based approach.
The central argument is that for normal functioning, either small
or large can be effective and adequate enough; but when the unexpected
happens, the results are more severe and widespread with larger systems. 

An issue specific to today\textquoteright s society, corruption, was
discussed. We considered why it would be useful to have organizations
that would nurture other smaller organizations. Lastly, we looked
at a way of life involving a quest for knowledge and applying it to
solve problems facing society, while being cognizant of the limitations
of knowledge. Such a way of life works best when organizations are
less complex.

The dynamic nature of any social science system means that the limited
predictive ability of any awareness will necessitate periodic reviews
and programs then need to be prescribed in response to what is required.
We certainly hope that this work will subsequently set the stage for
an investigative methodology using the four pronged approach.
\end{doublespace}

\begin{doublespace}

\section{Notes and References}
\end{doublespace}
\begin{enumerate}
\begin{doublespace}
\item \label{enu:Lego Blocks}Lego Blocks \href{https://en.wikipedia.org/wiki/Lego}{Lego Blocks: https://en.wikipedia.org/wiki/Lego}
\item \label{enu:B.S.-stands-for}B.S. stands for Bullshit, \href{ http://en.wikipedia.org/wiki/Bullshit}{BS:  http://en.wikipedia.org/wiki/Bullshit}
\item \label{enu:Sergey-Bubka}Sergey Bubka \href{http://en.wikipedia.org/wiki/Sergey_Bubka}{Sergey Bubka: http://en.wikipedia.org/wiki/Sergey\_{}Bubka}
\item Acs, Z. J., \& Varga, A. (2005). Entrepreneurship, agglomeration and
technological change. Small Business Economics, 24(3), 323-334. https://doi.org/10.1007/s11187-005-1998-4
\item Acs, Z. J., \& Varga, A. (2002). Geography, endogenous growth, and
innovation. International Regional Science Review, 25(1), 132-148.
https://doi.org/10.1177/016001702762039484
\item Anderson, T. R., \& Warkov, S. (1961). Organizational size and functional
complexity: a study of administration in hospitals. American Sociological
Review, 23-28. https://doi.org/10.2307/2090509
\item Ash, J. S., Berg, M., \& Coiera, E. (2004). Some unintended consequences
of information technology in health care. Journal of the American
Medical Informatics Association, 11(2).
\item Bacharach, S. B. (1989). Organizational theories: Some criteria for
evaluation. Academy of management review, 14(4), 496-515.
\item Bastiat, F. (1968). The law. Laissez Faire Books.
\item Beck, T. (2008). Bank competition and financial stability: friends
or foes?. World Bank Policy Research Working Paper Series. https://doi.org/10.1596/1813-9450-4656
\item Beck, T., \& Demirguc-Kunt, A. (2006). Small and medium-size enterprises:
Access to finance as a growth constraint. Journal of Banking \& Finance,
30(11), 2931-2943. https://doi.org/10.1016/j.jbankfin.2006.05.009
\item Bhargava, D. (1989). Manu Smriti: A Sociological Analysis. Rawat Publications.
\item Bockstette, V., Chanda, A., \& Putterman, L. (2002). States and markets:
The advantage of an early start. Journal of Economic Growth, 7(4),
347-369. https://doi.org/10.1023/A:1020827801137
\item Bühler, G. (Ed.). (1886). The laws of Manu (Vol. 25). Clarendon Press.
\item Busenitz, L. W., \& Barney, J. B. (1997). Differences between entrepreneurs
and managers in large organizations: Biases \& heuristics in decision-making.
Journal of business venturing. https://doi.org/10.1016/S0883-9026(96)00003-1
\item Carroll, A. B. (1991). The pyramid of corporate social responsibility:
toward the moral management of organizational stakeholders. Business
horizons, 34(4), 39-48. https://doi.org/10.1016/0007-6813(91)90005-G
\item Cornelius, W. A. (2001). Death at the border: Efficacy and unintended
consequences of US immigration control policy. Population and development
review, 27(4), 661-685. https://doi.org/10.1111/j.1728-4457.2001.00661.x
\item Creswell, J. W. (2013). Research design: Qualitative, quantitative,
and mixed methods approaches. Sage publications.
\item Daft, R. L. (1985). Why I recommended that your manuscript be rejected
and what you can do about it. Publishing in the organizational sciences,
193-209.
\item Damanpour, F. (1996). Organizational complexity and innovation: developing
and testing multiple contingency models. Management science, 42(5),
693-716. https://doi.org/10.1287/mnsc.42.5.693
\item De Nicolo, G. (2001, April). Size, charter value and risk in banking:
An international perspective. In EFA 2001 Barcelona Meetings.
\item Doidge, N. (2007). The brain that changes itself: stories of personal
triumph from the frontiers of brain science/Norman.
\item Diamond, J. (1999). Guns, germs, and steel: The fates of human societies.
WW Norton \& Company.
\item Friedman, M. (1970). The Social Responsibility of Business is to Increase
its Profits.
\item Grant, R. M. (1996). Toward a knowledge-based theory of the firm.
Strategic management journal, 17, 109-122. https://doi.org/10.1002/smj.4250171110
\item Hall, R. H., Johnson, N. J., \& Haas, J. E. (1967). Organizational
size, complexity, and formalization. American Sociological Review,
903-912. https://doi.org/10.2307/2092844
\item Hayek, F. A. (2009). The Road to Serfdom. University of Chicago Press.
\item Huxley, A. (2007). Brave new world. Ernst Klett Sprachen.
\item Kahneman, D., \& Krueger, A. B. (2006). Developments in the measurement
of subjective well-being. The journal of economic perspectives, 20(1),
3-24. https://doi.org/10.1257/089533006776526030
\item Kasarda, J. D. (1974). The structural implications of social system
size: A three-level analysis. American Sociological Review, 19-28.
https://doi.org/10.2307/2094273
\item Kashyap, R. (2010). Go Bald and Get Gold in Business. Working Paper.
https://doi.org/10.2139/ssrn.2338390
\item Kashyap, R. (2014a). Dynamic Multi-Factor Bid\textendash Offer Adjustment
Model. The Journal of Trading, 9(3), 42-55. https://doi.org/10.3905/jot.2014.9.3.042
\item Kashyap, R. (2015a). Financial Services, Economic Growth and Well-Being.
Indian Journal of Finance, 9(1), 9-22. https://doi.org/10.17010/ijf/2015/v9i1/71531
\item Kashyap, R. (2015b). A Tale of Two Consequences. The Journal of Trading,
10(4), 51-95. https://doi.org/10.3905/jot.2015.10.4.051
\item Kashyap, R. (2016). Hong Kong - Shanghai Connect / Hong Kong - Beijing
Disconnect (?), Scaling the Great Wall of Chinese Securities Trading
Costs. The Journal of Trading, 11(3), 81-134. https://doi.org/10.3905/jot.2016.11.3.081
\item Kashyap, R. (2016b). A Simple Model of Organizational Size and Complexity.
Working Paper.
\item Layard, R. (2010). Measuring subjective well-being. Science, 327(5965),
534-535. https://doi.org/10.1126/science.1186315
\item Linck, J. S., Netter, J. M., \& Yang, T. (2009). The effects and unintended
consequences of the Sarbanes-Oxley Act on the supply and demand for
directors. Review of Financial Studies, 22(8). https://doi.org/10.1093/rfs/hhn084
\item McDonald, P., \& Tang, Y. Y. (2014). Neuroscientific Insights Into
Management Development Theoretical Propositions and Practical Implications.
Group \& Organization Management. https://doi.org/10.1177/1059601114550712
\item Marx, K., \& Engels, F. (1959). The communist manifesto (Vol. 6008).
eBookEden. com.
\item Merton, R. K. (1936). The unanticipated consequences of purposive
social action. American sociological review, 1(6), 894-904. https://doi.org/10.2307/2084615
\item Nagel, E. (1961). The structure of science: Problems in the logic
of scientific explanation.
\item Norton, R. (2002). Unintended consequences. The Concise Encyclopedia
of Economics.
\item Paich, M., \& Sterman, J. D. (1993). Boom, bust, and failures to learn
in experimental markets. Management Science, 39(12), 1439-1458. https://doi.org/10.1287/mnsc.39.12.1439
\item Plato. (1945). The republic of Plato (Vol. 30, pp. 175-203). New York:
Oxford University Press.
\item Pollay, R. W. (1986). The distorted mirror: Reflections on the unintended
consequences of advertising. Journal of marketing, 50(2), 18-36. https://doi.org/10.2307/1251597
\item Porta, R. L., Lopez-De-Silane, F., Shleifer, A., \& Vishny, R. W.
(1996). Trust in large organizations (No. w5864). National Bureau
of Economic Research. https://doi.org/10.3386/w5864
\item Popper, K. R. (2002). The poverty of historicism. Psychology Press.
\item Popper, K. R. (1959). The logic of scientific discovery. London: Hutchinson.
\item Rizzo, J. R., House, R. J., \& Lirtzman, S. I. (1970). Role conflict
and ambiguity in complex organizations. Administrative science quarterly,
150-163. https://doi.org/10.2307/2391486
\item Sandel, M. J. (2010). Justice: What's the right thing to do?. Macmillan.
\item Schoorman, F. D. (1988). Escalation bias in performance appraisals:
An unintended consequence of supervisor participation in hiring decisions.
Journal of Applied Psychology, 73(1), 58. https://doi.org/10.1037/0021-9010.73.1.58
\item Seeley, T. D. (2009). The wisdom of the hive: the social physiology
of honey bee colonies. Harvard University Press.
\item Shleifer, A., \& Vishny, R. W. (1997). A survey of corporate governance.
The journal of finance, 52(2), 737-783. https://doi.org/10.1111/j.1540-6261.1997.tb04820.x
\item Stiglitz, J., Sen, A., \& Fitoussi, J. P. (2009). The measurement
of economic performance and social progress revisited. Reflections
and overview. Commission on the Measurement of Economic Performance
and Social Progress, Paris. Also, Commission Report (2010).
\item Sweeney, J., \& Sweeney, R. J. (1977). Monetary theory and the great
Capitol Hill Baby Sitting Co-op crisis: comment. Journal of Money,
Credit and Banking, 9(1), 86-89. https://doi.org/10.2307/1992001
\item Whetten, D. A. (1989). What constitutes a theoretical contribution?.
Academy of management review, 14(4), 490-495. https://doi.org/10.2307/258554
\item Winston, M. L. (1991). The biology of the honey bee. Harvard University
Press.\end{doublespace}
\end{enumerate}

\end{document}